\documentstyle[12pt]{article}

\newcommand{\be}{\begin{equation}}
\newcommand{\ee}{\end{equation}}
\newcommand{\bea}{\begin{eqnarray}}
\newcommand{\eea}{\end{eqnarray}}

\newlength{\del}
\newlength{\dlz}
\title{Critical exponents of the 3D antiferromagnetic \\
       three-state Potts model \\
       using the Coherent-Anomaly Method}
\author{Miroslav Kolesik\thanks{
        Permanent address:
        Institute of Physics, SAS, D\' ubravsk\' a cesta 9,
         Bratislava 842 28, Slovakia
        }\hspace {10pt}  and \hspace {10pt}
        Masuo Suzuki  \\
        Department of Physics, University of Tokyo, \\
          Bunkyo-ku, Tokyo 113, Japan}
\begin{document}
\maketitle

\begin{abstract}

The  antiferromagnetic  three-state  Potts  model on the  simple-cubic
lattice is studied using the  coherent-anomaly  method (CAM).  The CAM
analysis  provides the  estimates  for the  critical  exponents  which
indicate the XY universality  class, namely  $\alpha=-0.011 $, $\beta=
0.351  $,  $\gamma=  1.309 $ and  $\delta=  4.73 $.  This  observation
corroborates  the results of the recent Monte Carlo  simulations,  and
disagrees with the proposal of a new universality class.

\bigskip


\bigskip

\noindent
{\bf Running title:}  AF-Potts model critical exponents using CAM
\end{abstract}

\section{Introduction}

The  three-dimensional  antiferromagnetic  three-state Potts model has
been  investigated  intensively  during the last decade.  However, our
understanding  of its  low-temperature  properties  is still  far from
complete.  There  are   several   controversial   results   suggesting
different  universality  classes  of its phase  transition  as well as
different  types  of  the  low-temperature  ordering.  In the  present
article, we concentrate on the critical properties.

The existence of the finite-temperature  second-order phase transition
from  the  disordered  to  the  low-temperature   phase  is  generally
accepted,  but  there has been a  discussion  about  its  universality
class.  Let us review the previous results in brief.

Banavar  et al.  \cite{Banavar82}  conjectured  that the  model  under
investigation    belongs   to   the   universality    class   of   the
three-dimensional  XY ($O(2)$)  model.
 Ono    \cite{Ono}    observed    a
Kosterlitz-Thouless  phase with a  vanishing  order  parameter  in his
Monte Carlo  study, but that was probably a  consequence  of the short
simulation  time.  While  the  first  Monte  Carlo  estimates  of  the
critical exponents by Wang et al.  \cite{Wang89} lay somewhere between
the  Ising  and the XY  universality  classes,  improved  measurements
\cite{Wang90}  were in a good agreement  with the later.  On the other
hand, using the Monte Carlo twist  method,  Ueno et al.  \cite{Ueno89}
obtained the  critical  exponents  which  indicate a new  universality
class.  They  also  found  some   evidence   of  a  new  type  of  the
low-temperature  phase.  Later,  Okabe  \cite{Okabe92}  estimated  the
exponents  $\beta$  and  $\nu$ by Monte  Carlo  renormalization  group
approach,  and  obtained  results  which were again in favor of the XY
class.  Recently, Gottlob and Hasenbusch \cite{Gottlob94a, Gottlob94b}
obtained high-precision Monte Carlo estimates of $\gamma$ and $\nu$ in
a very good  agreement with the estimates for the XY model.  They also
measured various critical amplitudes, again in full agreement with the
XY universality  class.  Estimates for critical  indices are listed in
Table 1 together  with the  corresponding  values  calculated  for the
$O(2)$-model   \cite{LeGuillou80,   LeGuillou85}   .  It  seems   that
arguments  for the XY  universality  class  predominate.  On the other
hand,   Ueno   et   al.   \cite{Ueno93a}   later   argued   that   the
low-temperature  phase is an incompletely  ordered phase, and that its
nature  is not  compatible  with the XY  picture.  In order to help to
clear  up this  controversy,  we  present  our  results  based  on the
coherent-anomaly method.

We  use  the  same  technique  which  was  applied   recently  to  the
three-dimensional   Ising  model  and   provided   accurate   critical
exponents.  The   present   treatment   is  nothing   but  an  obvious
generalization   of  the  method  described  in  detail  in  the  Ref.
\cite{Kolesik94a}.  Therefore, we only briefly  sketch its main points
in the next  section.  In Section  3., we present  the  results of our
analysis.  Concluding remarks are given in Section 4.

\section{Mean-field solutions based on a series expansion}

The  coherent-anomaly  method  \cite{Suzuki86,  Suzuki87, Katori} is a
general  approach for  investigation  of critical  phenomena (see Ref.
\cite{Suzuki94}  for a recent review).  It is based on the analysis of
a suitable set of mean-field type approximations for the given system.
Here,    we   use   the    variational    series-expansion    approach
\cite{Kolesik93a,  Kolesik93b,  Kolesik93c,Kolesik94a} to generate the
series of approximate solutions.

Let  us  consider  the  simple-cubic  Potts  model  described  by  the
Hamiltonian
\be  {\cal  H}  =  \sum_{<i,j>}  \delta  (s_i  ,  s_j)  -
\sum_{i\in a}  \sum_{s=1}^3  H_s^a  \delta  (s,\  s_i)-
\sum_{i\in  b} \sum_{s=1}^3  H_s^b  \delta  (s,\  s_i)
\label{ham}  \ee
with spin variables  $s_i$ taking on three  different
values, say  $\{1,2,3\}$.  The first summation in (\ref{ham}) runs over
all nearest-neighbor  pairs of the cubic lattice, while the second and
the  third  summations  correspond  to the  interaction  with  external
fields.  We distinguish between the $a$- and  $b$-sublattices in order
to be able to take into account the antiferromagnetic  order appearing
below  the   critical   point.  The   notation   $h_s^{a,b}=\beta
H_s^{a,b}$ is used below for the dimensionless external fields.

The partition  function can be rewritten as follows:
\begin{eqnarray}
& Z =  \sum_{\{s_i\}}  \exp(-\beta  \cal H) =& \nonumber  \\
&\sum_{\{s_{abc}\}}  \prod_{(x  y z)}^{*}
w( s_{x y z}, s_{x+1 y z}, s_{x+1 y+1 z},
   s_{x y+1 z}, s_{x y z+1}, s_{x+1 y z+1},
   s_{x+1 y+1 z+1}, s_{x y+1 z+1} )&,
\label{suma}
\end{eqnarray}
where the product runs only over the triples $(x y z)$ in which all
entries are either even or odd, and
\newpage
\begin{eqnarray}
&w(s_1,s_2,s_3,s_4,s_5,s_6,s_7,s_8)  =  &\nonumber\\
  &\exp[-\beta(\delta(s_1,s_2)+\delta(s_2,s_3)+
               \delta(s_3,s_4)+\delta(s_4,s_1)+
               \delta(s_5,s_6)+\delta(s_6,s_7)+
                              \phantom{]}&\nonumber\\
  &\phantom{\exp[-\beta(}
               \delta(s_7,s_8)+\delta(s_8,s_5)+
               \delta(s_1,s_5)+\delta(s_2,s_6)+
               \delta(s_3,s_7)+\delta(s_4,s_8))]
                              \phantom{+}&\nonumber\\
&\times\exp[\sum_{s=1}^3 h_s^a
       (\delta(s,s_1)+\delta(s,s_3)+
        \delta(s,s_6)+\delta(s,s_8))]&\nonumber\\
&\times\exp[\sum_{s=1}^3 h_s^b
       (\delta(s,s_2)+\delta(s,s_4)+
        \delta(s,s_5)+\delta(s,s_7))]& .
\label{vahy}
\end{eqnarray}

The sum  (\ref{suma})  is  nothing  but the  partition  function  of a
three-state  vertex model  defined on the bcc lattice, with the vertex
weights $w$  determined by  (\ref{vahy}).  Our strategy is to make use
of the gauge  invariance  \cite{Wegner,  Gaaf} of the vertex models to
construct a set of mean-field solutions of the model.  We parameterize
the gauge transformation in the following way:
\begin{eqnarray}
&\tilde    w_a(s_1,s_2,s_3,s_4,s_5,s_6,s_7,s_8) =&\nonumber\\
&\sum_{\{r_i\}}
        A_{s_1,r_1}  B_{s_2,r_2}  A_{s_3,r_3} B_{s_4,r_4}
        A_{s_6,r_6}  B_{s_7,r_7}  A_{s_8,r_8} B_{s_5,r_5}
        w(r_1,r_2,r_3,r_4,r_5,r_6,r_7,r_8)  &
\end{eqnarray}
\begin{eqnarray}
&\tilde    w_b(s_1,s_2,s_3,s_4,s_5,s_6,s_7,s_8) =&\nonumber\\
&\sum_{\{r_i\}}
        B_{s_1,r_1}  A_{s_2,r_2}  B_{s_3,r_3}  A_{s_4,r_4}
        B_{s_6,r_6}  A_{s_7,r_7}  B_{s_8,r_8}  A_{s_5,r_5}
        w(r_1,r_2,r_3,r_4,r_5,r_6,r_7,r_8)  & ,
\end{eqnarray}
where  $A$ and $B$ are  orthogonal  matrices  with  their  first  rows
parameterized as
\begin{eqnarray}
&\{A_{11},A_{12},A_{13}\}=
    &\sqrt{1-x_1^2-x_2^2}\{1,1,1\}/\sqrt{3}  +
      x_1  \{0,1,-1\}/\sqrt{2}  +
      x_2  \{-2,1,1\}/\sqrt{6}  \nonumber \\
&\{B_{11},B_{12},B_{13}\}=
    &\sqrt{1-x_3^2-x_4^2}\{1,1,1\}/\sqrt{3}  +
      x_3  \{0,1,-1\}/\sqrt{2}  +
      x_4 \{-2,1,1\}/\sqrt{6}   \nonumber \\
\end{eqnarray}
Apart  from  the  orthogonality  condition,  the  remaining  rows  are
arbitrary;  their  parameterization  does not effect the  calculation.
Alternatively, we use the parameterization
\begin{eqnarray}
&x_1 = r_A \cos{\phi_A} &x_2 = r_A \sin{\phi_A}   \nonumber  \\
&x_3 = r_B \cos{\phi_B} &x_4 = r_B \sin{\phi_B}
\end{eqnarray}
when it is  appropriate.  The gauge  invariance  consists  in the fact
that the  partition  function  does not change when one  replaces  the
original  weights  $w$ by the  transformed  weights  $\tilde  w_a$ and
$\tilde  w_b$ on the  sublattices  (of the bcc  lattice)  $a$ and $b$,
respectively.

Following  the usual  procedure  of the  variational  series-expansion
method, we  generate a formal  series  expansion  for the  transformed
vertex model  described by the weights  $\tilde w_a$ and $\tilde  w_b$
without  fixing  the gauge  parameters  $\{x_i\}$.  There are  totally
$2\times  3^8$ kinds of the vertex  weights for a general  three-state
model  with two  nonequivalent  sublattices.  For the  purpose  of the
series   expansion,  we  classify  them  into   $2\times  495$  classes
$\{\omega_i^{a,b}\}_{i=0}^{494}$  induced by the lattice symmetry.  We
fix   the    notation    such    that    $\omega_0^{a,b}    =   \tilde
w_{a,b}(1,1,1,1,1,1,1,1)$.  Then,  we  calculate  the  formal   series
expansion for the free energy in powers of $\omega_i^a/\omega_0^a$ and
$\omega_i^b/\omega_0^b$.  (See  Refs.  \cite{Kolesik94a,   Kolesik94b}
for technical  details.)  Thus, our expansion for the logarithm of the
partition function has the form
\be
{\cal F}_L ={1\over 2}\log(\omega_0^a \omega_0^b) +
          \sum_{n=2}^L f_n( \{ \omega_i^{a,b}/\omega_0^{a,b} \})\ ,
\ee
where $\{f_n\}$ are homogeneous  polynomials of order $n$, and $L$ denotes
the  maximal  order  included in the  expansion.  Because of the large
number  of  variables  $\omega$,  this  is a huge  formula  containing
thousands of terms and we cannot present it here \footnote{The  series
will be available  upon  request  from  fyzikomi@savba.savba.sk}.  The
calculation of the series is the limiting  factor in the variational
series-expansion approach.
In this case we were able to  generate  the  series  only up to the order
$L=6$.   Nevertheless,    within   the   present    formulation,   the
approximation   ${\cal   F}_6$   includes   quite   large   excitation
encompassing  up to 40 original spins.  (Note that the  graph-counting
here is rather different from that of the usual series expansions).

Having  calculated  the formal  expansion  for the free energy, we can
return to the original model described by the weights $\tilde w_a$ and
$\tilde   w_b$.  Thus,   ${\cal  F}_L$   becomes  a  function  of  the
temperature,  external fields and of the gauge  parameters  $\{x_i\}$.
Within  the  variational series-expansion
method,  the  gauge  parameters  are  fixed  by  the
stationarity conditions
\be
{\partial {\cal F}_L \over \partial x_i} = 0 \quad (i=1,2,3\ {\rm and}\ 4).
\label{cond}
\ee
Let us describe the  structure of the  solutions to  (\ref{cond}).  In
the  high-temperature  phase and in zero external  fields, there exists
only  a  single  solution,  namely  $x_i=1$  (i.e.  $r_A=r_B=0$).  New
solutions  characterized  by finite  $r_A$  and  $r_B$  appear  at the
critical  point.  However,  our  mean-field   solutions  ${\cal  F}_L$
exhibit a nearly perfect rotational symmetry in a broad region around the
critical  point.  This  means  that  we have  always  $r_A=r_B=r$  and
$\phi_A=\phi_B+\pi=\phi$,  and, moreover,  ${\cal  F}_L(r,\phi)$  {\em
does  not  depend}  on  the  angle   $\phi$.  Perturbations   of  this
rotational symmetry are at least of the order $r^6$, and are therefore
completely {\em irrelevant} to the CAM analysis.  This property can be
shown explicitly in the lowest-order approximation.

The symmetric  properties of our solutions  reflect the restoration of
the rotational symmetry of the model at its critical point, and are in
agreement with what was observed in Ref.  \cite{Gottlob94a}.  In fact,
without this  property, it would be impossible to extract any reliable
estimates for critical exponents using the CAM.

\section{CAM analysis}

The  coherent-anomaly  method is based on the scaling of the so-called
mean-field critical  coefficients  \cite{Suzuki86,  Suzuki87, Katori}.
Similarly  as in  Ref.\cite{Kolesik93b},  we  expand  the  approximant
${\cal  F}_L$ in the vicinity of its critical  point
in order to calculate  the
coefficients.  The  only  difference  from  the  calculation  in  Ref.
\cite{Kolesik93b}  is that we have four gauge  parameters to take into
account.  We are  interested  mainly  in the  magnetizations  and  the
suscetibilties
\be
m_i^\alpha = {\partial {\cal F} \over \partial h_i^\alpha } \, \ \ \ \
\chi_{ij}^{\alpha \beta} =
       {\partial m_i^\alpha \over \partial H_j^\beta }.
\ee
Naturally,  these quantities are not  independent.  Actually, there is
only one independent susceptibility because we have $\chi_{ii}^{\alpha
\alpha}   =    -\chi_{kl}^{\alpha    \alpha}/2   \   (k\ne   l)$   and
$\chi_{kl}^{\alpha \beta} = -\chi_{kl}^{\alpha  \alpha}$.  In the same
way, we  have  effectively  only  one  critical  coefficient  for  the
magnetizations  because  of the  rotational symmetry  mentioned  above.
Therefore, we omit the indices $i$ and $j$  corresponding to the three
Potts states as well as the sublattice indices $\alpha$ and $\beta$.

After  some  straightforward  calculations,  we  obtain  the  following
expressions for the mean-field  critical  coefficients of the specific
heat $\bar  c_L$, the  magnetization  $\bar  m_L$, the  susceptibility
$\bar \chi_L$ and the critical magnetization $\bar m^c_L$ :
\begin{eqnarray}
\bar c_L &=&\beta^c_L \beta^* (\partial_{rr\beta}{\cal F})^2 /
            \partial_{rrrr}{\cal F}
            \label{cecko} ,\\
\bar m_L &=&
             \partial_{hx}{\cal F}
             (-6 \beta^c_L \partial_{rr\beta} {\cal F}/
             \partial_{rrrr}{\cal F} )^{1/2}  ,\\
\bar \chi_L &=&
             \partial_{hx}{\cal F} \partial_{hx}{\cal F} /
             \partial_{rr\beta} {\cal F}\\
{\rm and}\ \ & & \nonumber \\
\bar m^c_L &=&
             \partial_{hx}{\cal F}
             (-6 \beta^c_L \partial_{hx} {\cal F}/
             \partial_{rrrr}{\cal F} )^{1/3} .
             \label{decko}
\end{eqnarray}
Here, the symbol $\partial_x$ means the derivative with respect to the
gauge parameter $x_1$ or $x_2$, $\partial_h$ stands for the derivative
with respect to the external field, and $\partial_r$ is the derivative
with respect to the radius gauge  parameter.  All the  derivatives  in
(\ref{cecko}-\ref{decko})  are to be calculated at the critical  point
$\beta=\beta^c_L$,  $x_i=1$  ($r=r_A=r_B=0$).  Taking  different gauge
parameters $x$'s and external fields $h$ results in equivalent sets of
mean-field critical  coefficients.  Namely, the critical  coefficients
can be rescaled so that they may be equal to unity in the lowest order
approximation  $L=0$.  This does not affect the  subsequent  analysis,
and it turns out that one is left  with  only one set of  coefficients
for each quantity.  This is a consequence of the model symmetry.

Now, we can estimate the true critical exponents of the model from the
CAM scaling formulas as \cite{Kolesik94a}
\begin{eqnarray}
\bar c_L  & \sim &
\left({t^c_L}\over{t^*}\right)^{\alpha /2}
\left({\vert t^*-t^c_L\vert }\over{t^*}\right)^{-\alpha}
= \left(\vert \Delta_L \vert\right)^{-\alpha},
\label{cczac}
\\
\bar m_L  & \sim &
\left({t^c_L}\over{t^*}\right)^{(1/2-\beta)/2}
\left({\vert t^*-t^c_L\vert }\over{t^*}\right)^{\beta -1/2}
= \left(\vert \Delta_L \vert\right)^{\beta -1/2},
\\
\bar\chi_L  & \sim  &
\left({t^c_L}\over{t^*}\right)^{(\gamma-1)/2}
\left({\vert t^*-t^c_L\vert }\over{t^*}\right)^{1-\gamma}
= \left(\vert \Delta_L \vert\right)^{1-\gamma}
\\
{\rm and}\ \ & & \nonumber
\\
\bar m^c_L  & \sim &
\left({t^c_L}\over{t^*}\right)^{\psi /2}
\left({\vert t^*-t^c_L\vert }\over{t^*}\right)^{-\psi}
= \left(\vert \Delta_L \vert\right)^{-\psi} ,\;
  \psi=\gamma (\delta-3)/3(\delta-1)
\label{cckon}
\end{eqnarray}
where $\Delta_L=(t^*/t^c_L)^{1/2}-(t^c_L/t^*)^{1/2}$ with $t$ standing
for the temperature or for the inverse temperature; $t^*$ is the exact
critical value while $t_L^c$ is its $L$ th order approximation.

In order to extract  accurate  estimates  of critical  exponents, it is
necessary to know the critical  temperature  with a high accuracy.  We
have used the value  $\beta^*=0.81563$
obtained from the high-precision  Monte
Carlo  simulation  by Gottlob and  Hasenbusch  \cite{Gottlob94a},  and
fitted the critical  exponents to the above  coherent-anomaly  scaling
formulas.  We would like to stress  that within the CAM  approach  the
resulting    exponents   always   fulfill   the   scaling    relations
$\alpha+2\beta+\gamma=2$ and $\gamma=\beta(\delta-1)$.

We excluded the approximants for $L=0,2$ and $3$
from our analysis as usual,
because their critical behavior is the same.  Actually, they represent
the same approximation  which has a Bethe-like  character; it does not
take into account  properly  even the  shortest  cycle of the lattice.
This is why the critical coefficients for $0\le L \le 3$ do not follow
the CAM scaling very well for the type of the expansion used here.  On
the other hand, even the  approximation  with $L=4$  takes  account of
quite large  loops, and  reflects the  structure  of the lattice  much
better.

Therefore, we have fitted the critical  exponents to the data obtained
only from the approximations  with $L=4,5$ and $6$.
The values thus obtained are
shown in the  bottom of Table 2, and the  corresponding  CAM plots are
presented in Fig.  1.  Unfortunately, unlike in the Ising case, we are
not  able to  estimate  the  error  of our  results,  because  we have
effectively  only three data points.  However,  the same method  works
extremely well for the Ising model, which is in fact a special case of
the present  system.  This is why we believe that our estimates  would
be essentially  unchanged also for higher-order  approximations.  Now,
let us compare  our  results to the  previously  calculated  exponents
shown  in  Table 1.  Our  estimates  of  $\gamma$  are in a very  good
agreement  with  the the values of the XY  universality  class.
The  values  for
$\alpha$  and $\beta$  are also  consistent  with the XY  universality
class,  though  the  differences   between  various  fits  indicate  a
relatively larger error.

\section{ Conclusion}

We have  calculated  the  estimates  of critical  exponents  $\alpha$,
$\beta$,    $\gamma$    and    $\delta$   of   the    thre-dimensional
antiferromagnetic  Potts  model.  Our  calculation  is  based  on  the
variational-expansion  approach  combined  with  the  coherent-anomaly
method.  The most  difficult  part of the  computation  in the present
investigation  was the  generation  of the series  expansion.  We have
calculated   the  necessary   series  up  to  the  order  $L=6$.  This
relatively  short series provides only three points for the subsequent
CAM analysis, and therefore, we cannot estimate the error bars for our
estimates.  The extension of the series seems to be rather  infeasible
even  for the  order  $L=7$.  Hence,  the  present  method  may not be
competitive  with the  Monte  Carlo  simulation  for the  model  under
investigation.  However,   all  the  recent   estimates   of  critical
exponents  calculated  directly  for the Potts  model  come from Monte
Carlo  studies.  In  the  present situation in which  similar  results
by  different methods are not available,
it is important to obtain at least {\em some} results independently of
the Monte Carlo studies, in order to confirm the universality class.
The
agreement of our estimates with the XY universality  class is fairly
good and
the discrepancy between our results and the values proposed by Ueno is
large.  We thereby  conclude that our  estimates  corroborate  the XY
universality class, and are clearly against the new universality class
as proposed by Ueno \cite{Ueno89}.

Finally, let us make a brief  remark  concerning  the  low-temperature
phase.  As  described   above,  our   mean-field   solutions   exhibit
approximate  rotational symmetry in the  order-parameter  space.  From
the  point  of view  of the  mean-field  critical  behavior, we can
regard them as
perfectly  symmetric,  because the deviations from the symmetry do not
affect  the  calculation  of  the  mean-field  critical  coefficients.
However, it is the asymmetry  which  determines  the  direction of the
sublattice-symmetry  breaking.  Unfortunately, the deviations from the
perfect  symmetry are so small that the  differences  between the free
energies  corresponding  to  different  solutions under the  stationarity
conditions  (\ref{cond})  are  practically  negligible  compared to the
expected  accuracy  of  our  approximations.  This  is why  we  cannot
determine the type of the  low-temperature  ordering  from the present
calculations.  It is  possible  that  this  peculiar  behavior  of our
approximations  reflects  the  nature  of the  low-temperature  phase,
namely  that the order  parameter  of the  model  becomes  effectively
continuous.  This  would  be in  line  with  the  restoration  of  the
rotational symmetry    observed    by    Gottlob    and    Hasenbusch
\cite{Gottlob94a}.  It  would  also  elucidate  the  origin  of the XY
critical   indices.  Clearly,  the   interesting   properties  of  the
low-temperature phase deserve further investigation.

\bigskip\bigskip
\noindent{\large\bf Acknowledgment}

\bigskip
One of us (M.K.) would like to express his gratitude to the Nishina
Memorial Foundation for granting him a scholarship.

\vfill
\newpage
\newcommand{\rf}[5]{{\rm #1, }{ #2 }{ #3} (#4) #5. }

\vfill
\newpage
\centerline{\large\bf Table 1.}
\bigskip
\begin{center}
\begin{tabular}{|l|c|l|l|l|l|l|} \hline
\rule[-0.25cm]{0cm}{0.7cm} Author & year & Ref. &\hfil $\alpha$ &\hfil
                 $\beta$ &\hfil  $\gamma$ &\hfil  $\nu$ \\ \hline
Wang & 1989 & \cite{Wang89} &\rule{0cm}{\dlz} & & 1.27(5) & 0.63(4)\\
Wang & 1990 & \cite{Wang90} & & & 1.31(3) & 0.66(3)\\
Ueno & 1989 & \cite{Ueno89} & 0.15 & 0.34(2) & 1.10(2) & 0.58(1) \\
Okabe& 1992 & \cite{Okabe92}& & 0.33(2) & & 0.66(2) \\
Gottlob & 1994 & \cite{Gottlob94a} & & & 1.310(9) & 0.664(4)\\
Gottlob & 1994 & \cite{Gottlob94b} & & &  & 0.663(4)\\ \hline
LeGuillou & 1980 & \cite{LeGuillou80} & \rule{0cm}{\dlz} $-0.007(6)$
                                    & 0.345(2) & 1.3160(25) & 0.669(2)\\
LeGuillou & 1985 & \cite{LeGuillou85} &\rule[-0.25cm]{0cm}{0.7cm}
                                    & 0.3485(35) & 1.315(7) & 0.671(5)\\ \hline
\end{tabular}
\end{center}
\vfill

\noindent {\bf Table 1.}  Recent  estimates for critical  exponents of
the antiferromagnetic three-state Potts model in three dimensions.  We
have also included the  exponents  calculated  for the 3D XY model for
comparison.   The    values    in    Refs.   \cite{LeGuillou80}    and
\cite{LeGuillou85}  are the standard RG estimates and the results from
the $\epsilon$-expansion, respectively.

\newpage
\centerline{\large\bf Table 2.}
\bigskip
\begin{center}
\begin{tabular}
{|l|c|c|c|c|c| }
\hline\rule[-0.25cm]{0cm}{0.7cm}
\rule{0cm}{\dlz}$L$ & $T^c_L$ & $\bar c_L$ & $\bar m_L$ &
$\bar\chi_L$ & $\bar m_L$  \\
\hline
\rule{0cm}{\dlz} 0-3 & 1.3943  & 1.0\phantom{0000} & 1.0\phantom{0000} &
1.0\phantom{0000} & 1.0\phantom{0000}  \\
\rule{0cm}{1pt}  4 & 1.3014  & 1.09957  & 1.15608  & 1.21551  & 1.17556 \\
\rule{0cm}{1pt}  5 & 1.2998  & 1.11207  & 1.16766  & 1.22603  & 1.18680 \\
\rule[-0.3cm]{0pt}{5pt} 6 & 1.2735 & 1.09982 & 1.23997 & 1.39797  & 1.29054 \\
\hline
\multicolumn{2}{|l|}{\rule{0cm}{\dlz}  data used \rule{0cm}{1pt}}
 & $\alpha$ & $\beta$ & $\gamma$ & $\delta$ \\
\multicolumn{2}{|l|}{\rule{0cm}{\dlz}5,6:}
 & $-0.025$ & 0.359 & 1.306 &  4.63 \\
\multicolumn{2}{|l|}{\rule{0cm}{0cm}4,6:}
 & $\phantom{-}0.001$ & 0.344 & 1.311 &  4.81 \\
\multicolumn{2}{|l|}{\rule{0cm}{1pt}4,5,6:}
 & $-0.011$ & 0.351 & 1.309 &  4.73 \\
\hline
\end{tabular}
\end{center}
\vfill

\noindent   {\bf  Table  2.}  Critical   temperatures   and   critical
mean-field coefficients as calculated for various approximation orders
$L$.  The   critical   exponents   were   fitted   to  the  data  from
approximations with $L>3$ and the resulting estimates are shown in the
bottom row of the table.  The numbers in the first  column  correspond
to the approximations used.

\newpage
\begin{Large}
\noindent {\bf Figure caption}
\end{Large}

\bigskip
{\bf Fig.  1.}  CAM scaling of the critical mean-field coefficients $\bar Q_L$
for the specific heat, $\bar c_L$ ($\Diamond$), magnetization, $\bar m_L$
($+$), susceptibility, $\bar \chi_L$ ($\Box$) and for the critical
magnetization, $\bar m^c_L$ ($\times$).  The distance from the true
critical point is measured in $\Delta_L =
(\beta^*/\beta^c_L)^{1/2}-(\beta^c_L/\beta^*)^{1/2}$.  Critical
coefficients were rescaled so that $\bar Q_0$=1 (see the text).

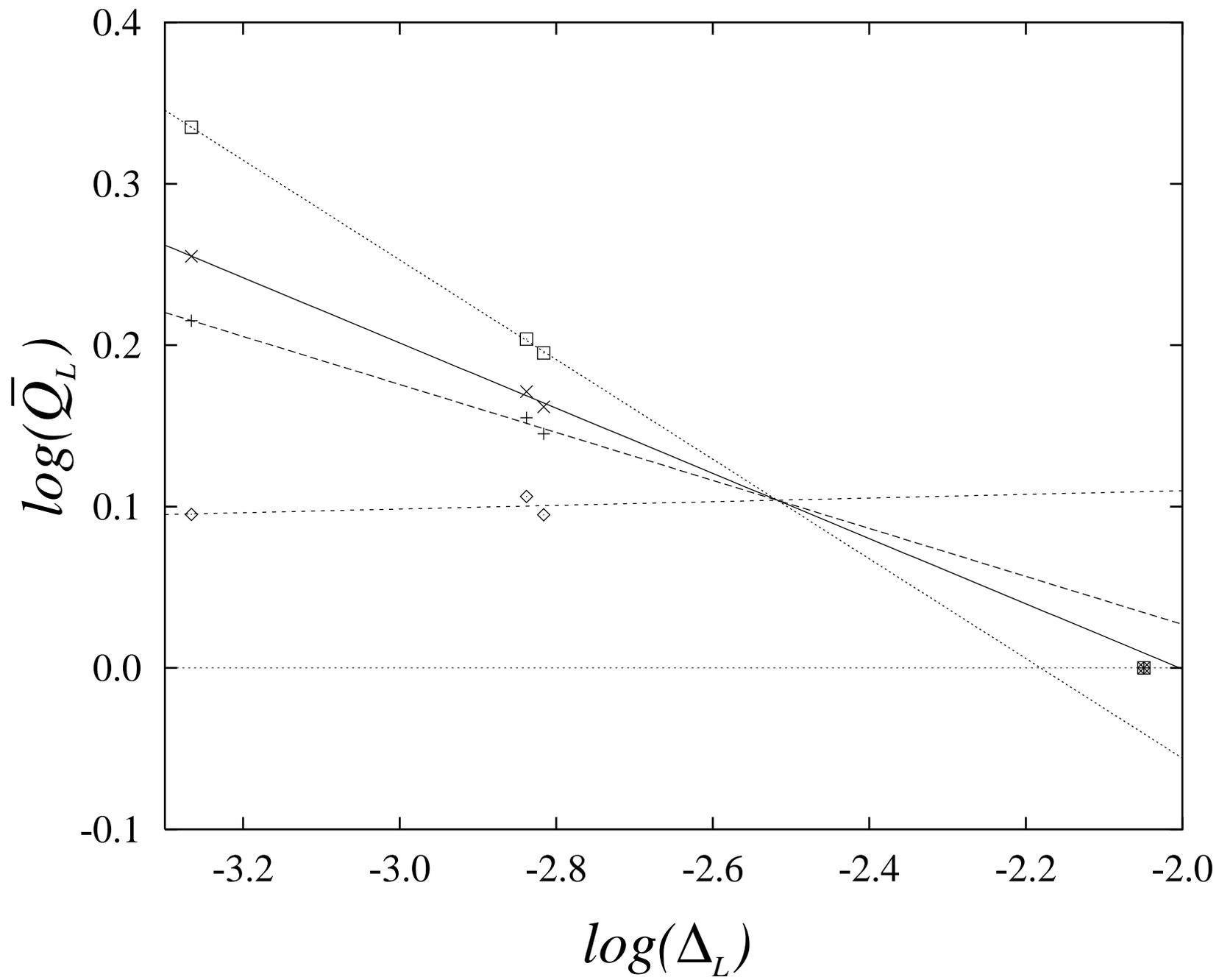
\end{document}